\documentclass[conference]{IEEEtran}

\usepackage{amsthm}
\usepackage[cmex10]{amsmath}
\usepackage{amsfonts}
\usepackage{amssymb}
\usepackage{cite}

\ifCLASSINFOpdf
\usepackage[pdftex]{graphicx}
\usepackage{subfigure}
\else

\usepackage{graphicx}
\usepackage{subfigure}
\fi
\usepackage{epstopdf}
\usepackage{color,xcolor}
\usepackage{multicol}  
\usepackage{multirow}
\usepackage{booktabs} 
\usepackage{threeparttable}
\usepackage{algorithmic,algorithm}
\usepackage{subfigure}

\usepackage[font=normalsize]{caption}

\usepackage{algorithmic,algorithm}

\usepackage[numbers,sort&compress]{natbib}
\usepackage{color}

\usepackage{url}

\usepackage{mdwtab}

\begin{document}

\title{ByteSGAN: A Semi-supervised Generative Adversarial Network for Encrypted Traffic Classification of SDN Edge Gateway in Green Communication Network\\
	\thanks{The paper is sponsored by National Natural Science Fundation (General Program) Grant 61972211, China, and National Key Research and Development Project Grant 2020YFB1804700, China.}
}

\author{\IEEEauthorblockN{1\textsuperscript{st} Pan Wang}
	\IEEEauthorblockA{\textit{School of Modern Posts} \\
		\textit{Nanjing University of Posts \& Telecommunications}\\
		Nanjing, China \\
		wangpan@njupt.edu.cn}
	\and
	\IEEEauthorblockN{2\textsuperscript{nd} Zixuan Wang}
	\IEEEauthorblockA{\textit{School of Modern Posts} \\
		\textit{Nanjing University of Posts \& Telecommunications}\\
		Nanjing, China \\
		wangzx@runtrend.com.cn}
	\and
	\IEEEauthorblockN{3\textsuperscript{rd} Feng Ye}
	\IEEEauthorblockA{\textit{Department of Electrical \& Computer Engineering} \\
		\textit{University of Dayton}\\
		Dayton, OH, USA \\
		fye001@udayton.edu}
	\and
	\IEEEauthorblockN{4\textsuperscript{th} Xuejiao Chen}
	\IEEEauthorblockA{\textit{Department of Communication} \\
		\textit{Nanjing College of Information Technology}\\
		Nanjing, China \\
		chenxj@njcit.cn}
}

\maketitle
	
\begin{abstract}
With the rapid development of Green Communication Network, the types and quantity of network traffic data are accordingly increasing. Network traffic classification become a non-trivial research task in the area of network management and security, which not only help to improve the fine-grained network resource allocation, but also enable policy-driven network management. Meanwhile, the combination of SDN and Edge Computing can leverage both SDN at its global visiability of network-wide and Edge Computing at its low latency and good privacy-preserving. However, capturing large labeled datasets is a cumbersome and time-consuming manual labor. Semi-Supervised learning is an appropriate technique to overcome this problem. With that in mind, we proposed a Generative Adversarial Network (GAN)-based Semi-Supervised Learning Encrypted Traffic Classification method called \emph{ByteSGAN} embedded in SDN Edge Gateway to achieve the goal of traffic classification in a fine-grained manner to further improve network resource utilization. ByteSGAN can only use a small number of labeled traffic samples and a large number of unlabeled samples to achieve a good performance of traffic classification by modifying the structure and loss function of the regular GAN discriminator network in a semi-supervised learning way. Based on public dataset 'ISCX2012 VPN-nonVPN', two experimental results show that the ByteSGAN can efficiently improve the performance of traffic classifier and outperform the other supervised learning method like CNN. 

\end{abstract}


\section{Introduction}\label{sec:intro} 
With the rapid development of Green Communication Network, the types and quantity of network traffic data are accordingly increasing. Network traffic classification become a non-trivial research task in the area of network management and security. It is undoubtedly the cornerstone of dynamic access control, network resources scheduling, QoS provisioning, intrusion and malware detection etc. High efficient and accurate traffic classification is of great practical significance to provide service quality assurance, dynamic access control and abnormal network behaviors detection. Nonetheless, with the widespread adoption of encryption techniques for network including Green Communications, 5G and IoT, the growth of portion of encrypted traffic has dramatically arisen a huge  challenge for network management like QoS provisioning. Therefore, accurate and robust encrypted traffic classification not only help to improve the fine-grained network resource allocation, but also enable policy-driven network management.

Software defined networking(SDN) is envisioned as an emerging and promising networking paradigm with promise to dramatically improve network resource utilization, simplify network management, reduce operating costs and promote innovation and evolution. Meanwhile, Edge Computing with its inherent nature of low latency, high bandwidth and good privacy-preserving has attracted a lot of attentions of both acdemic and industry. Intuitively, some researchers started to think of how to push the intelligence to the edge of Green Communications Network and exploit the benefit of SDN. Thus, the combination of SDN and Edge Computing  give rise to SDN Edge Gateway (SDN-EGW), which can leverage both SDN at its global visiability of network-wide,programmability for traffic control and Edge Computing at its low latency and good privacy-preserving.

Howerver, the state-of-the-art solutions usually adopt payload-based methods like Deep Packet Inspection (DPI) or flow statistics based  methods like Machine-Learing (ML) for encrypted traffic classification. Nevertheless, DPI technology incurs high computational costs and requires frequent manual signature maintenance, morever, nowadays DPI techniques are always prone to error, time-consuming and costly due to payload encryption and privacy issues. Although ML can alleviate some limitations of DPI to some extent, its poor accuracy and handcrafted feature selection by manual labor still can not meet the fine-grained QoS-awareness requirements of Green Communication Network. Unlike most traditional ML algorithms, Deep Learning (DL) performs automatic feature extraction without human intervention, which undoubtedly makes it a highly desirable approach for traffic classification, especially encrypted traffic.  Recent research work has demonstrated the superiority of DL methods in traffic classification~\cite{MobileTC-2018}, such as MLP~\cite{Datanet}, CNN~\cite{deeppacket,Wang-1D-CNN,Wang2D-CNN,Seq2Img,HierarchicalTC}, SAE~\cite{blackhat}, LSTM~\cite{IoT-CNN-2017,HAST-IDS}. 

As above mentioned, previous research works mostly require large quantites of labeled datasets, namely, supervised learning, which has been the center of most researching in DL. However, as we all know, capturing large labeled datasets is a cumbersome and time-consuming manual labor. Hence, the necessity of creating models capable of learning from fewer data is increasing faster. With that in mind, semi-supervised learning is an appropriate technique to overcome this problem. Additionally, labeling traffic datasets is non-trivial and cumbersome process. There are two labeling methods for traffic datasets, one is DPI based and the other is scripts based labeling which can mimic human behavior, respectively. Whereas, DPI based labeling lose efficacy because of traffic encryption and scripts based methods are not always accurate. In contrast to these limitations of labeling data, unlabeled data is always abundant and easily-obtainable. Hence, How to use semi-supervised learning to leverage the readily available unlabeled traffic data for accurate traffic classification appears to be particularly important.

In this paper, we proposed a Generative Adversarial Network (GAN)-based Semi-Supervised Learning Encrypted Traffic Classification method called \emph{ByteSGAN} embedded in SDN Edge Gateway to achieve the goal of traffic classification in a fine-grained manner to further improve network resource utilization. ByteSGAN can only use a small number of labeled traffic samples and a large number of unlabeled samples to achieve a good performance of traffic classification by modifying the structure and loss function of the regular GAN discriminator network in a semi-supervised learning way. Based on public dataset 'ISCX2012 VPN-nonVPN', two experimental results show that the ByteSGAN can efficiently improve the performance of traffic classifier and outperform the other supervised learning method like CNN.

The rest of this paper is organized as follows. Section II introduces preliminaries and related works of traffic classification and SDN Edge Gateway . Section III describes the principles of GAN and SGAN. Section IV illustrates the methodology of ByteSGAN. The experimental results are provided and discussed in Section V. Section VI  concludes our work.

\section{Preliminaries}\label{sec:related_works}
\subsection{ML and DL based approach of Traffic Classification}
ML based classification methods always take payload-independant parameters as model input, such as packet length, inter-arrival time and flow duration to circumvent the problems of encrypted content and user's privacy~\cite{2007legal}. There are plethora of works focusing on ML algorithms during the last decades. In general, there are two learning strategies used: one is the supervised methods like decision tree, SVM and Naive Bayes, the other is unsupervised approaches like k-means and PCA~\cite{2008NguyenTC}. Nevertheless,  many drawbacks hindered ML based methods widely applied to traffic classification, such as handcrafted traffic features driven by domain-expert, time-consuming, unsuited to automation, rapidly outdated when compared to the evolution. Unlike most traditional ML algorithms, Deep Learning performs automatic feature extraction without human intervention, which undoubtedly makes it a highly desirable approach for traffic classification, especially encrypted traffic.  Recent research work has demonstrated the superiority of DL methods in traffic classification~\cite{MobileTC-2018,deeppacket,Datanet,Wang-1D-CNN,Wang2D-CNN,IoT-CNN-2017,HAST-IDS,Seq2Img,blackhat,TC-VAE,HierarchicalTC}. The workflow of DL based classfication usually consists of three steps, which has been demonstrated in our previous work~\cite{mobileTC-survey}. First, model inputs are defined and designed according to some principles, such as raw packets, PCAP files or flow statistics features. Second, models and algorithms are elaborately chosen according to models' characteristics and aim of the classifier. Finally, the DL classifier is trained to automatically extract the features of traffic.  

\subsection{Traffic Classification of Edge Gateway in SDN}

\begin{figure}[ht!]
	\centering\includegraphics[width=3.4 in]{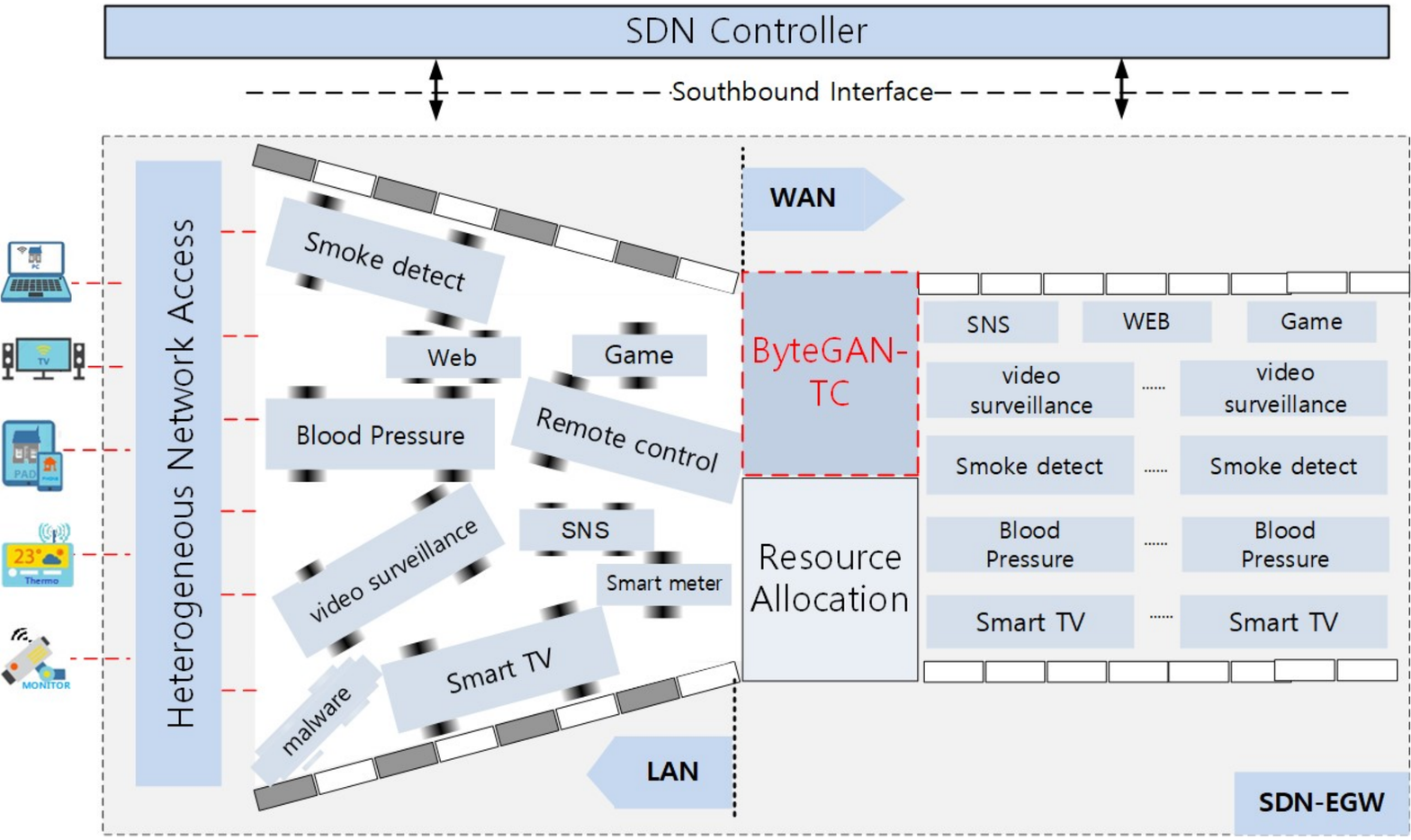} 
	\caption{The scheme of traffic classification over SDN-EGW.}\label{fig:Fig_framework} 
\end{figure}

In this subsection, we will clarify the necessity and range of our research - \emph{ByteSGAN}. As shown in Fig.~\ref{fig:Fig_framework}, several kinds of smart devices (Blood pressure monitors/video surveillance/Smart TV, etc.) connect to the SDN-EGW via hetergeneous access technology (WiFi/ZigBee/Bluetooth, etc.). However, all packets from those smart devices will queue on WAN interface and wait to be forwarded by SDN-EGW out-of-order. Therefore, you can see three aspects about research necessity as follows:
\begin{enumerate}
	\item There are different QoS requirements in smart home. Current typical Green Communications Networks are composed of several heterogeneous IoT applications, such as Home Automation, Healthcare and Entertainment appliances, in which tens of devices need different QoS policy. For example, healthcare of acute conditions like blood pressure monitoring need real-time and high priority but it is low rate, in contrast, healthcare of long-term observations need relatively low priority and non-real-time but it maybe needs high bandwidth consumption like high definition  medical images. 
	
	\item No QoS-Aware capability in current Green Communications Network. In legacy network, QoS-Aware capability depends on additional dedicated equipments like Traffic Management devices based on traffic shaping technology because TCP/IP can not ganrantee the QoS under the limitation of "best effort" forward mechanism. In contrast, although SDN can manage and control traffic flows by means of central control and global view of the whole network, it still has not extended its capability to the edge, even home network.
	
	\item No fine-grained traffic classification capability in SDN-EGW, especially encrypted traffic classification. There are some previous research work about SDN traffic classification~\cite{MultiClassifier2014,application_aware2017,SDN_DPI2017}, however, most of them focus on traffic classification over SDN controller. It is apparent that SDN controller will suffer great traffic processing pressure. Very few research works about traffic classification over SDN-HGW have proposed some methods using DPI or ML~\cite{SDN_DPI2017,SDN_DPI_APP2017,SDN_ML_QOS2016,SDN_ML2017}. However, these methods can not handle encrypted traffic accurately and in a fine-grained way. With the rapid growth of encrypted application, such as e-commerce, searching and SNS, more and more applications will be delivered via end-to-end encrypted channels, such as HTTPS and SRTP, therefore, encrypted traffic classification is facing a very severe challenge.

\end{enumerate}

In order to overcome above mentioned shortcomings of Green Communication Network, We inevitably will face to three important challenges: First, traffic classification over encrypted traffic; Second, the incorporation of traffic classification capability and SDN-HGW; Third, resource allocation based on traffic classification and SDN facilities (SDN Controller and vRGW). In this paper, we will focus on the first problem.

\section{Generative Adversarial Networks (GAN)}\label{gan}
GAN provide a promising way to learn deep representation without extensively annotated training data. They achieve this through deriving backpropagation signals through a competitive process involving a pair of networks. GAN can be applied in a variety of applications like image synthesis, style transfer, semantic image editing, which are gaining more and more attention from both academic and industry.
\subsection{GAN}\label{AA}
\begin{figure}[htbp]
	\centerline{\includegraphics[width=8cm, height=6cm]{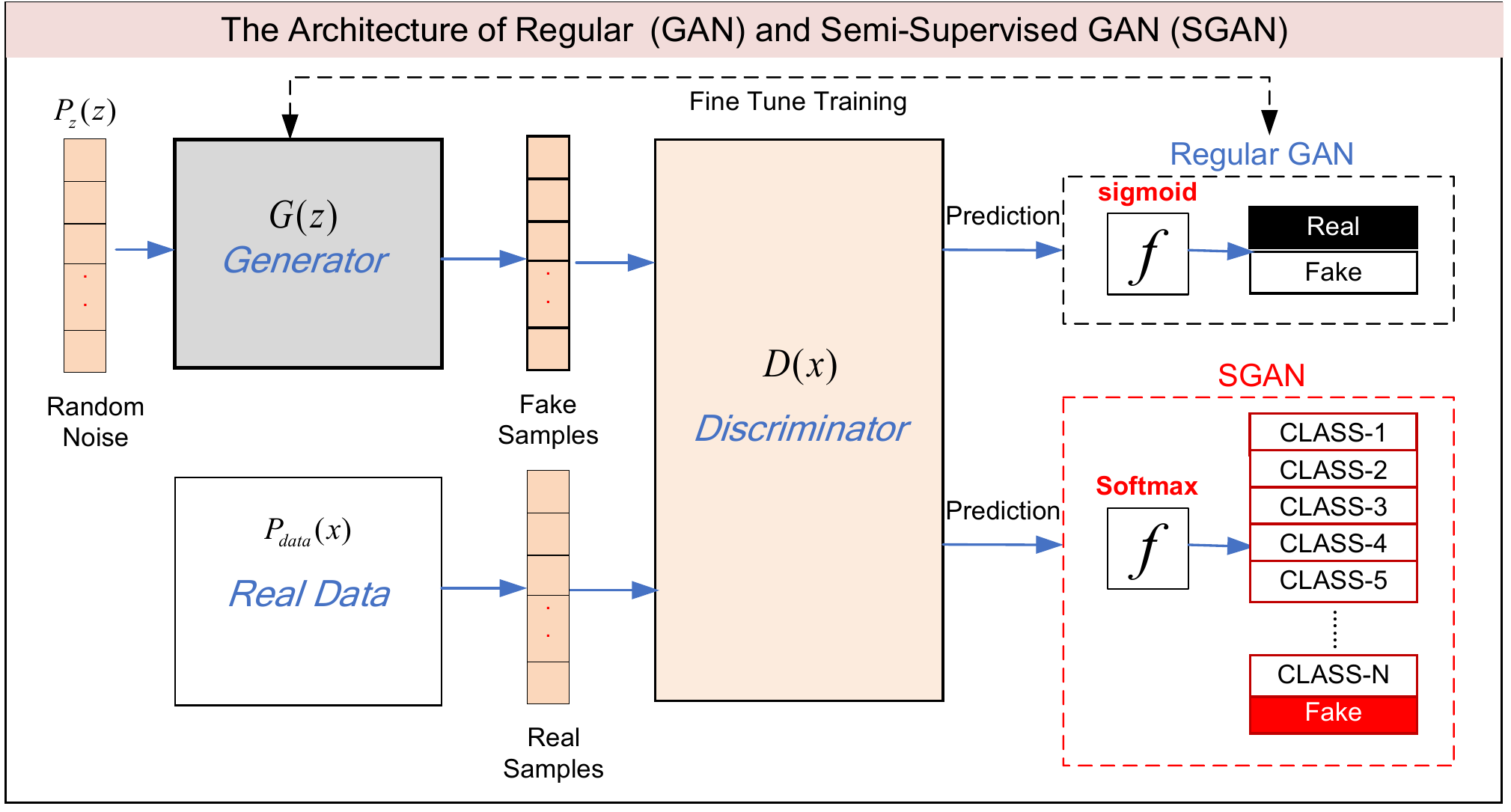}}
	\caption{The Basci Architecture of GAN.}
	\label{fig_gan}
\end{figure}

A regular GAN consists of two parts, the generator $G$ and the discriminator $D$. As shown in Fig.~\ref{fig_gan},
The role of the generator is to take random noise as input by learning the characteristic distribution of real data. The discriminator aims at determining whether the data is real or generated by $G$.  
The generator $G$ simulates the feature distribution $P_g$ of the real data by the prior distribution $P_z(z)$. The input of the discriminator is the real and generated data, correspondingly, the output $D(x)$ indicates the probability of whether the input data is real or not~\cite{r11}. During the training process, $G$ and $D$ play a two-player mini-max game until $D$ can't judge whether the sample data is real, which means that the two networks reach the Nash Equilibrium. The objective function of GAN can be expressed by \eqref{GAN}:
\begin{equation}
\begin{split}
\label{GAN}
\mathop {\min }\limits_{\rm{G}} \mathop {\max }\limits_D V(D,G) = \mathbb{E}_{x\sim{p_{data}}(x)}[\log D(x)] \\
+ {\mathbb{E}_{z\sim{p_z}(z)}}[\log (1 - D(G(z)))]
\end{split}
\end{equation}

In Equation \eqref{GAN}, $P_{data}(x)$ represents distribution of the real data. When training $D$, the goal is to optimize the probability of TRUE $D(G(z))$ as small as possible and the probability of TRUE $D\left(x\right)$ of the real data $x$ as much as possible. When training $G$, the goal is to make $D(G(z))$ as much as possible. From \eqref{GAN}, we can calculate the optimal discriminator as \eqref{eq1}. As can be seen from \eqref{eq1} below, when $P_{data}\left(x\right)=P_z\left(z\right)$, it means that $D$ cannot distinguish whether the sample is true or false, $D$ and $G$ reach the Nash Equilibrium, and the discriminator output is 0.5.

\begin{equation}
\begin{split}
\label{eq1}
D(x)=\frac{P_{data}(x)}{P_{data}(x)+P_z(z)}
\end{split}
\end{equation}

\subsection{Semi-Supervised GAN (SGAN)}
Semi-Supervised GAN, so called SGAN is an extension of GAN to the semi-supervised context by forcing the discriminator network to output class labels. A generative model $G$ and a discriminator $D$ are trained on a dataset with inputs belonging to one of N classes. At training time, $D$ is made to predict which of N+1 classes the input belongs to, where an extra class is added to corresponding to the outputs of $G$. The study of \cite{odena2016semisupervised} showed that this method can be used to create a more data-efficient classifier than a regular GAN.

As shown in Fig.~\ref{fig_gan}, the traditional discriminator network $D$ always outputs an estimated probability that the input samples is fake. This process  typically implemented with a feed-forward network ending in a sigmoid unit, however, if we implement this with softmax output layer with one unit for each of the classes [REAL, FAKE], apparently $D$ could have N+1 output units corresponding to [CLASS-1, CLASS-2,...CLASS-N, FAKE]. In this case, $D$ can also act as classifier, namely $C$. We can call this network $D/C$. While training an SGAN, it is similar to a GAN except that $D/C$ is trained to minimize the negative log likelihood with regard to the given labels and $G$ is trained to maximize it.

\section{The Methodology of ByteSGAN Encrypted Traffic Classification}\label{algorithm}
As we all know, obtaining large labeled datasets is a cumbersome and time-consuming manual labor. In contrast to these limitations of labeling data, unlabeled data is always abundant and easily-obtainable. Apparently, it is particularly important that how to use semi-supervised learning to leverage the readily available unlabeled traffic data for accurate traffic classification. Therefore, intuitively, how to combine the capability of the GAN-based encrypted traffic generation and semi-supervised learning for classification tasks become a very innovative research task.

\subsection{The Workflow of ByteSGAN}\label{algorithm_description}

\begin{figure}[ht!]
	\centering\includegraphics[width=3.3 in]{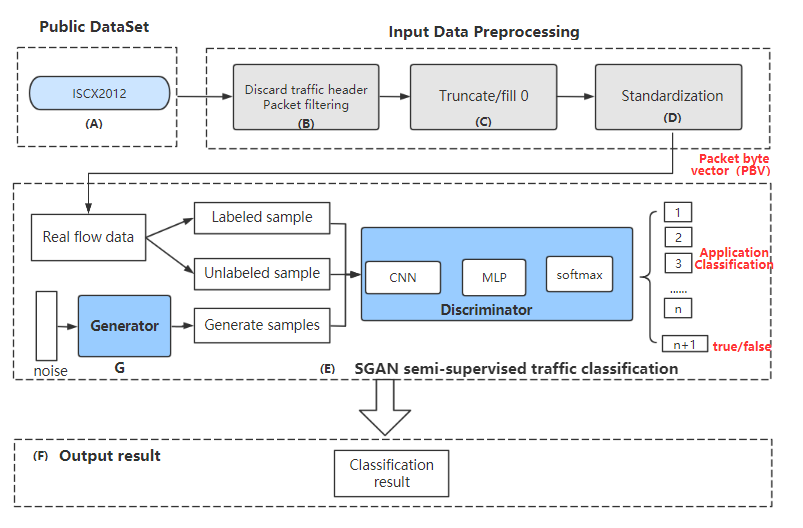} 
	\caption{The Workflow of ByteSGAN.}\label{fig:SGAN_Architecture} 
\end{figure}

As shown in Fig.~\ref{fig:SGAN_Architecture}, ByteSGAN mainly includes two steps: Input data pre-processing and SGAN based traffic classification.
First, we have to filter some packet data in the public dataset like ARP/DHCP/ICMP which is not relevant to specific applications. In addition, packet data will be truncated or padding with zero in order to normalize to a Packet Byte Vector (PBV)~\cite{Datanet}. Then the standardized PBV (i.e. real flow data) will be sent to ByteSGAN, namely SGAN semi-supervised traffic classification model as model input. It is worth noting that the labeled and unlabeled data in PBV are both fed into the discriminator network $D$ as real data. Then the generator network $G$ and discriminator network $D$ are alternately trained and iteratively updated multiple times to converge and output the classification results.

In ByteSGAN, multiple traffic samples can be classified and trained at the same time, which effectively solves the problem that the regular or original GAN has to be trained separately for each application. Moreover,  we can easily overcome the limitation of the labeled data by means of Semi-Supervised Learning.

\subsection{Loss function}\label{algorithm_lossfunction}
The discriminator $D$ of ByteSGAN is an N+1 dimensional classifier. Its input is a packet data sample and output is an N+1 dimensional vector, which can be represented as class probability, where formula  \eqref{x_false} means the $x$ is false, instead, formula  \eqref{x_ture} means the $x$  is true as well as it belongs to the class $i$.

\begin{equation}
	\begin{split}
		\label{x_false}
		P_{\text {model }}(y=N+1 \mid x)=\frac{\exp \left(c_{N+1}\right)}{\sum_{j=1}^{N+1} \exp \left(c_{j}\right)}
	\end{split}
\end{equation}

\begin{equation}
	\begin{split}
		\label{x_ture}
		P_{\text {model }}(y=i \mid x, i<N+1)=\frac{\exp \left(c_{\mathrm{i}}\right)}{\sum_{j=1}^{N+1} \exp \left(c_{j}\right)}
	\end{split}
\end{equation}

\subsection{Training process }\label{algorithm_lossfunction}
The training of ByteSGAN uses a combination of supervised and unsupervised way, which can significantly improve the learning ability of the model. Specifically, during the semi-supervised training, the discriminator $D$ and the generator $G$ are trained alternately and the parameters are updated iteratively. On one hand, while training the generator, feature learning will be used to make the generated traffic as close to the real data distribution as possible. On the other hand, while training the discriminator, the network parameters are iteratively updated to minimize the cross-entropy of the labeled traffic sample and the predicted probability distribution. The network model parameters of unlabeled samples and generated samples are updated by adversarial training. 
As show in Alg.\ref{algorithm:SGAN}, the training steps are as follows:
\begin{enumerate}

	\item Generate a random vector with Gaussian noise $z$ and input it to the generator network $G$, then get the generated traffic $G(z)$;

	\item Input both the generated traffic $G(z)$ and the labeled\&unlabeled real traffic $x$ into the discriminator network $D$ in batches, then output the sum of probability of $D(x)$ and $D[G(z)]$ by the activation function softmax;

	\item Fix the parameters of the generator network $G$. If the input flow is generated flow (i.e. fake data), we can use $E_{x \sim G} \log \left[p_{\text {model }}(y=N+1 \mid x)\right]$ as the loss function; if the real flow $x$  has been labeled,  we can use $L_{\text {labeled }}$  as the loss function; if the real flow $x$ is not labeled, we can use $L_{\text {unlabeled }}$ as the loss function. Adjust the parameters of the discriminator $D$ by Adam Gradient Descent method;

	\item Fix the parameters of the discriminator network $D$, perform feature matching operations on the real traffic data $x$ and the generated traffic $G(z)$ , then select the output of the fully connected layer as the middle layer and use feature matching to adjust the parameters of the generator network $G$;

	\item Repeat steps (1)-(4) until the number of iterations is completed.
\end{enumerate}

\begin{algorithm}[htbp]
	\caption{ The model training algorithm of ByteSGAN}  \label{algorithm:SGAN}
	\begin{algorithmic}[1]  
		\REQUIRE Real labeled sample $X$ after data preprocessing
		\ENSURE Accuracy of traffic classification
		\FOR {number of training iterations }
		\FOR { $k$ steps }
		\STATE Take a minibatch from the $m$ noise samples $\left\{z^{(1)}, \ldots, z^{(m)}\right\}$  whose noise prior distribution is $\mathrm{p}_{\text {g}}(z)$ .
		\STATE Take a minibatch from the $m$ training samples $\left\{x^{(1)}, \ldots, x^{(m)}\right\}$  with data distribution $\mathrm{p}_{\text {data}}(x)$.
		\STATE If the real traffic is not marked, use $L_{\text {unlabeled }}$  as a loss function:
		\STATE Update the discriminator through Adam gradient descent.
		\STATE If the real traffic $x$ has been marked, use $L_{\text {labeled }}$ as a loss function:
		\STATE Update the discriminator through Adam gradient descent:
		\STATE $\nabla_{\theta_{d}} \frac{1}{m} \sum_{i=1}^{m}\left[-\log \left[p_{\text {model }}\left(y=i \mid x^{(i)}, i<N+1\right)\right]\right]$
		\STATE If the input flow is generated flow, use $L_{\text {g}}$ as a loss function:
		\STATE Update the discriminator through Adam gradient descent:
		\STATE $\nabla_{\theta_{d}} \frac{1}{m} \sum_{i=1}^{m}\left[\log \left[p_{\text {model }}\left(y=N+1 \mid x^{(i)}\right)\right]\right]$
		\ENDFOR \\ 
		\STATE Take a minibatch from $m$ noise $p_{z}(z)$ samples whose prior distribution of noise is $\left\{z^{(1)}, \ldots, z^{(m)}\right\}$ .    
		\STATE Update generator.
			
		\ENDFOR \\ 
	\end{algorithmic}  
\end{algorithm}  

\subsection{Network Structure of ByteSGAN}\label{Network structure}
We refered to the network structure of DCGAN ~\cite{DCGAN} and the definition of semi-supervised model in the paper of Tim Salimans et al. ~\cite{ImprovedTrainingGANS} to adjust the input and output, the network structure of generator and discriminator  of ByteSGAN to make model training more stable. The detailed adjustments are as follows:

\begin{enumerate}
	\item The convolutional layer with strides is used instead of the pooling layer.
	\item The discriminator network $D$ adopts convolution instead of the pooling layer and the generator network $G$ model employs deconvolution instead of the pooling layer.
	\item LeakyReLU ~\cite{LeakyReLU} is used for activation in both the generator $G$ and the discriminator $D$. In addition, the generator $G$'s output layer adopts the tanh as the activation function according to the paper~\cite{ImprovedTrainingGANS}. \item The output layer of the discriminator $D$ is built as a stacked model with shared weights. First, build a supervised model by using $k$ classes of outputs and a softmax activation function. Then define the unsupervised model, which accepts the output of the supervised model before the softmax activation function, and then calculates the normalized sum of the exponential output.
\end{enumerate}

\begin{equation}
	\begin{split}
		\label{network_structure}
		D(x)=\frac{Z(x)}{Z(x)+1}, \text { where } Z(x)=\sum_{k=1}^{K} \exp \left[l_{k}(x)\right]
	\end{split}
\end{equation}

The output of the unsupervised model before the softmax activation function are all very small positive or negative values. When using formula \eqref{network_structure} , we will get the output of (0.0-1.0), which means that the model is encouraged to output strong predictions for real samples and small predictions or low activations for fake samples.

\subsection{The Generator Model Structure}\label{Generator structure}
\begin{figure}[ht!]
	\centering\includegraphics[width=3.3 in]{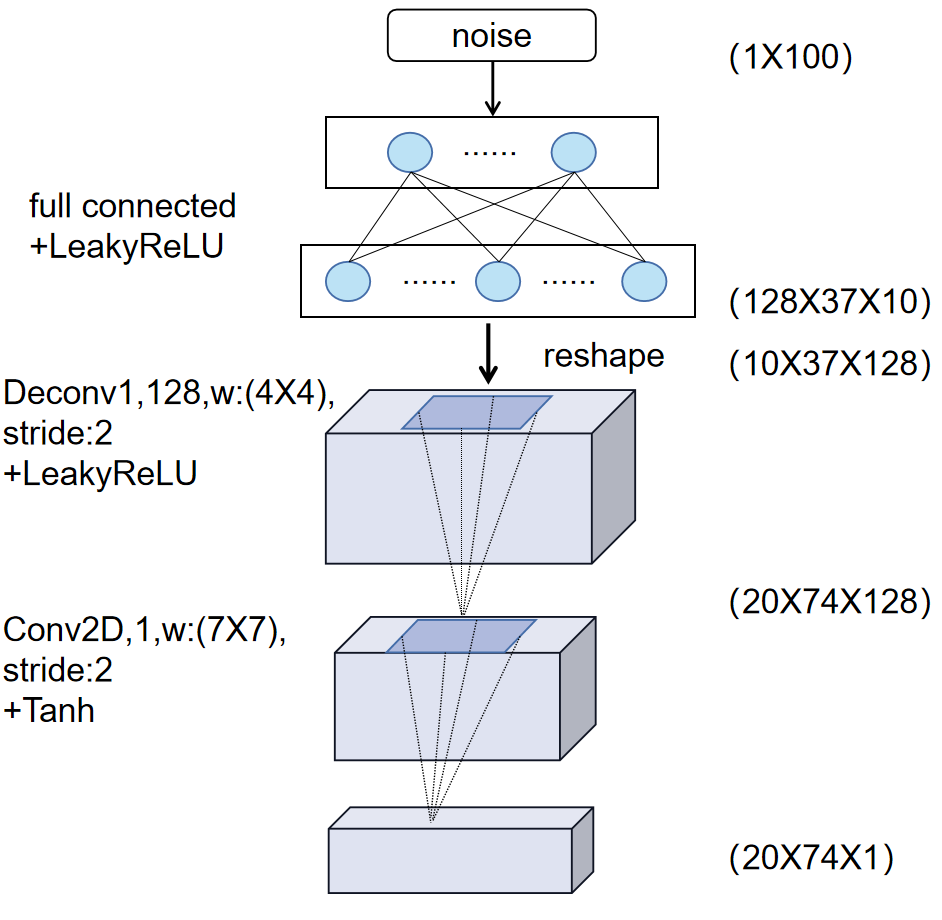} 
	\caption{The Structure of Generative Network Model }\label{fig:SGAN generation network structure diagram} 
\end{figure}

The network parameters are shown in  Fig.~\ref{fig:SGAN generation network structure diagram} . The generator $G$ is constructed by a one layer deconvolution network and a one layer convolution network. Firstly, we will input a 100-dimensional random noise $z$ with the Gaussian distribution into the fully connected network and make it into a three-dimensional tensor through dimensional transformation (i.e. reshape). Input the dimensionally transformed tensor into the convolution kernel $w$ with a size of 4*4 and stride of 2, which is activated by LeakyReLU and then input to a convolution layer $z$ with a kernel size of 7*7 and stride of 2, using Tanh activation to generate a traffic sample tensor of (20*74*1).

\subsection{The Discriminator Model Structure}\label{Discriminator structure}
\begin{figure}[ht!]
	\centering\includegraphics[width=3.3 in]{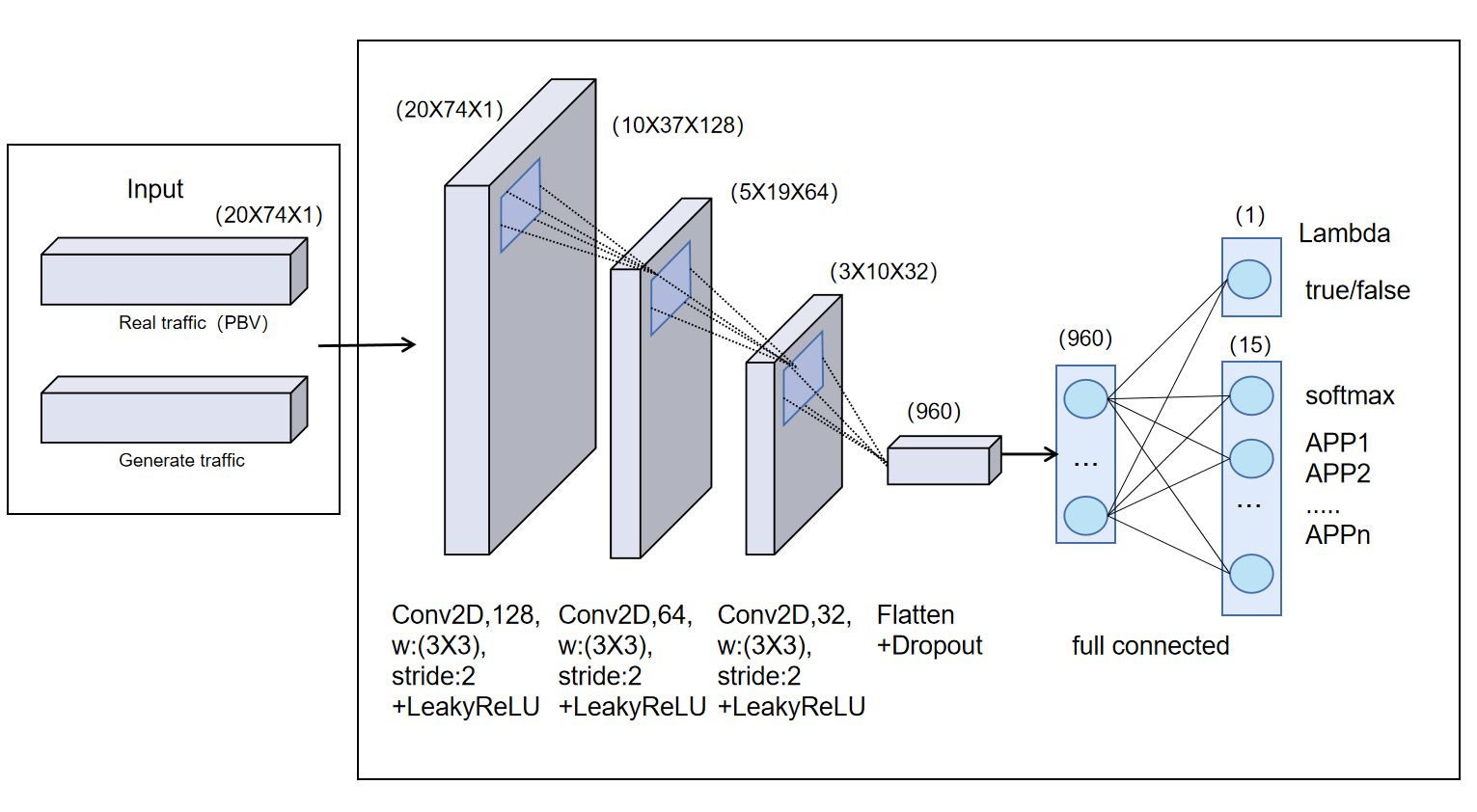} 
	\caption{The Structure of Discriminator Network Model}\label{fig:SGAN discriminant network structure diagram} 
\end{figure}

As shown in Fig.~\ref{fig:SGAN discriminant network structure diagram}, the discriminator $D$ is composed of 3 convolutional layers and 2 fully connected layers. The real traffic sample PBV with a length of 1480 after input data preprocessing is transformed to a three-dimensional tensor (20*74*1) by dimensional transformation (i.e. reshape), then sent to a 3-layer convolution kernel $w$ with a size of 3* 3. LeakyReLU is used for activation after each convolution. LeakyReLU can retain a small slope in the negative half axis (in this paper it is set to 0.2). Compared with the ReLU activation function, LeakyReLU can avoid the problem of the gradient vanish during training. After flattening through the Flatten layer, it is input to the fully connected network which is built as a stacked network with shared weights and adopts Lambda as activation function, and then exploys softmax to output normalized categories probabilities.

\section{Evaluation and Experimental results }\label{exp:results}
\subsection {Experimental Settings}\label{exp:setting}
\subsubsection{Dataset for Evaluation}\label{exp:dataset}
The experiments in this chapter are based on the "ISCX VPN-nonVPN traffic data set" as shown in Table~\ref{tab:Desc_Samples_ISCX}. Two experiments are carried out:
\begin{enumerate}
	\item In the first experiment, we use the same number of labeled samples and different numbers of unlabeled samples to perform classification experiments on 7 applications to verify the impact of the number of unlabeled samples on the ByteSGAN classification results.
	
 	\item In the second experiment, we use the same dataset with different number of labeled samples  to perform classification experiments on 15 applications by ByteSGAN and CNN, respectively in order to verify the improvement of classification accuracy of ByteSGAN with small number of labeled samples.
\end{enumerate}

\begin{table}[htbp]	
	\centering  
	\fontsize{6.5}{8}\selectfont  
	\begin{threeparttable}  
		\caption{Description of the chosen datasets from ISCX.}  \label{tab:Desc_Samples_ISCX}  
		\begin{tabular}{l|c|cc|cc|}  
			\toprule  
			\multirow{2}{*}{\textbf{Application}}&
			\multirow{2}{*}{\textbf{Security}}&
			\multicolumn{2}{c}{\textbf{Dataset samples}} \cr  
			\cmidrule(lr){3-4} \cmidrule(lr){5-6}  
			&\textbf{Protocol} & \textbf{Quantity} & \textbf{Percentage}\cr  
			
			\midrule  
			AIM				&HTTPS	 	   &4869&2.356\%\cr  
			Email-Client&SSL			  &4417&2.137\%\cr  
			Facebook	&HTTPS	 	   &5527&2.674\%\cr  
			Gmail		   &HTTPS		  &7329&3.546\%\cr  
			Hangout		&HTTPS	 	   &7587&3.671\%\cr  
			ICQ				&HTTPS	 	   &4243&2.053\%\cr  
			Netflix			&HTTPS		   &51932&25.126\%	\cr  
			SCP				&SSH	  		 &15390&7.446\%\cr  
			SFTP			&SSH	 		 &4729&2.287\%\cr  
			Skype		   &proprietary	 &4607&2.229\%\cr  
			Spotify		   &proprietary	 &14442&6.987\%\cr  
			torTwitter	  &proprietary	&14654&7.089\%\cr  
			Vimeo		  &HTTPS		  &18755&9.074\%\cr  
			voipbuster	&proprietary   &35469&17.161\%\cr  
			Youtube		&HTTPS			 &12738&6.163\%\cr  							
			\midrule
			
			TOTAL&  &{\bf 206688}&{\bf 100\%}
		\end{tabular}  
	\end{threeparttable}  
\end{table}  

\subsubsection{Configurations of the Computing Platform}\label{exp:config}
The experimental environmental parameters of this paper are shown in Table~\ref{tab:parameters}. The performance evaluations are conducted using a Dell R730 server with an Intel I7-7600U CPU 2.8 GHz, 16 GB RAM and an external GPU (Nvidia GeForce GTX 1050TI). The software platform for deep learning is built on Keras library with Tensorflow (GPU-based version 1.13.1) as the back-end support.

\linespread{1.5}
\begin{table}[htbp]
	\caption{Experimental Environment Parameters}
	\begin{center}
		\begin{tabular}{c c}
			\hline
			\textbf{Category}&{\textbf{Parameters}} \\
			\hline
			GPU & Nvidia GPU(GeForce GTX 1050Ti)  \\
			Operating System & Win 10 \\
			Deep learning platform & TensorFlow 1.13.1 + Keras 1.0.7\\
			CUDA Version & 9.0\\
			CuDNN Version & 7.6.0\\
			\hline
		\end{tabular}
		\label{tab:parameters}
	\end{center}
\end{table}

\subsection {The first experiment}\label{exp:setting}
In order to verify the impact of unlabeled data on the experimental results, it is necessary to ensure that there are enough data in the dataset. Therefore, seven application samples with more than 10,000 data are selected, and 10,000 of them are randomly selected as shown in Table ~\ref{tab:Experimental One Dataset}.

\begin{table}[htbp]
	\caption{Dataset of the first experiment }  \label{tab:Experimental One Dataset}
	\fontsize{6.5}{8}\selectfont  
	\begin{center}
		\begin{tabular}{c| c| c}
			\hline
			\textbf{Application Name)}&{\textbf{Quantity}} &{\textbf{Proportion)}} \\
			\hline
			Netflix		&10000&14.29\%	\cr  
			SCP				&10000&14.29\%\cr  
			Spotify		&10000&14.29\%\cr  
			torTwitter	&10000&14.29\%\cr  
			Vimeo		  &10000&14.29\%\cr  
			voipbuster &10000&14.29\%\cr  
			Youtube    &10000&14.29\%\cr  
			Total      &10000&100.00\%\cr
			\hline
		\end{tabular}
	\end{center}
\end{table}

On the selected data set, 1000 labeled samples are used at the same time, and 4000, 6000, and 8000 unlabeled samples are used for three experiments. The stochastic gradient descent method is used for training, and the batch parameter is 256. The noise is randomly sampled from the uniform distribution of [-1, 1], and the sample size is 100. The learning rate parameter is 0.0002, and the Adam optimizer is used to optimize the loss function of the generator and the discriminator.
The results of the three experiments are shown in Table~\ref{tab:Experimental One Results}:
\begin{table}[htbp]
	\caption{The first experimental results}  \label{tab:Experimental One Results}
	\fontsize{6.5}{8}\selectfont  
	\begin{center}
		\begin{tabular}{c| c| c}
			\hline
			\textbf{Number of labeled samples}&{\textbf{Number of unlabeled samples (per class)}} &{\textbf{Accuracy)}} \\
			\hline
			1000		&4000&98.21\%	\cr  
			1000		&6000&98.96\%\cr  
			1000		&8000&99.18\%\cr  
			\hline
		\end{tabular}
	\end{center}
\end{table}
The experimental results show that increasing the number of unlabeled samples can improve the classification accuracy of ByteSGAN when the number of labeled samples is constant.

\subsection {The second experiment}\label{exp:setting}
\subsubsection{ Experimental description}\label{exp:config}
\begin{figure}[ht!]
	\centering\includegraphics[width=3.3 in]{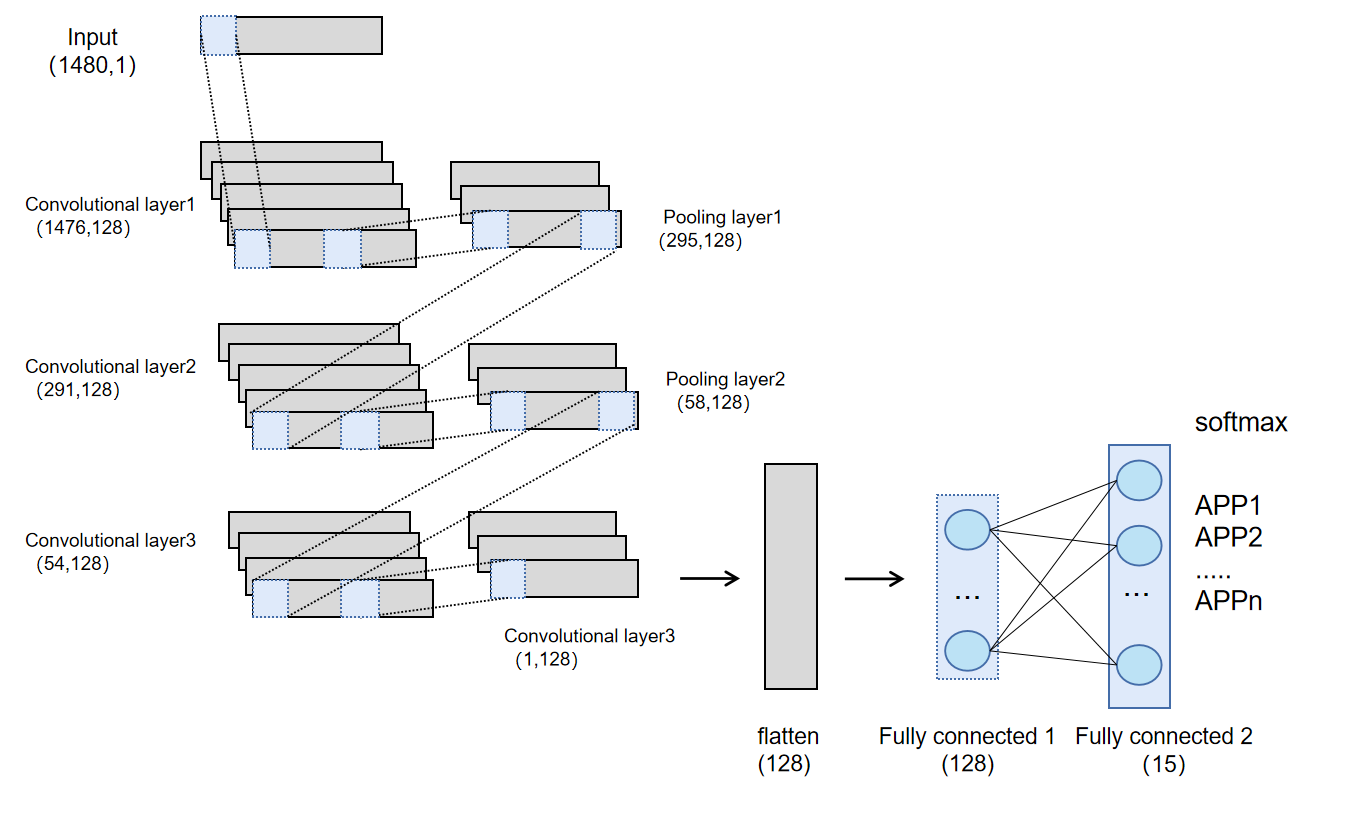} 
	\caption{CNN based traffic classification model structure}\label{fig:CNN_structure} 
\end{figure}

In order to compare with the classification experiment, the data set as shown in Table~\ref{tab:Desc_Samples_ISCX}  was used to conduct experiments on both the CNN-based and the ByteSGAN-based classification model. Before the experiment, select a specified number of labeled data from the dataset and the rest as unlabeled data for the experiment.
We adopt the stochastic gradient descent method for training and the batch parameter is 256. The noise is randomly sampled from the uniform distribution of [-1, 1], and the sample size is 100. The learning rate parameter is 0.0002, and the Adam optimizer is used to optimize the loss function of the generator and the discriminator.
For the sake of comparison, a simple CNN based classification model is designed and implemented and its network structure is shown in Fig.~\ref{fig:CNN_structure}. 
The adopted CNN-based encrypted traffic recognition model includes an input layer, three convolution and pooling connection layers, and an output layer. The input layer is packet data after the preprocessing. The specific parameters of the convolution kernel pooling layer are shown in the figure. The output layer adopts softmax classification. The neurons of output layer represent 15 different applications. The detailed model structure is as follows:
\begin{enumerate}
	\item The input layer data adopts the Packet Byte Vector (PBV) with (1480, 1) as input.
	\item The convolution layer in the first layer performs convolution operation uses 128 filters with a size of 5 and a step size of 1. The output result is calculated by the activation function ReLU (Shaped Linear Unit).
	\item Max-pooling is adopted with a size of 5 and stride of 1.
	\item The convolution layer in the second layer of convolution and pooling layer uses 128 filters with a size of 5 and a stride of 1. The output result is calculated by ReLU.
	\item The parameters used by the pooling layer in the second layer are the same as those in step (3).
	\item The convolution layer in the third layer uses 128 filters with a size of 5 and a stride of 1. The output result is calculated by ReLU.
	\item The pooling layer in the third layer employs maxi-pooling, with a size of 35 and a stride of 1.
	\item Through two fully connected layers, the network is transformed into a one-dimensional vector for subsequent classification.
	\item The output layer uses softmax for classification with 15 output units corresponding to 15 different encrypted traffic applications

\end{enumerate}

On the selected dataset, a small batch stochastic gradient descent method is used for training, and the batch parameter is 128. Use Rmsprop optimizer for the cross entropy loss function.
In this paper, we use ByteSGAN to perform classification experiments under the condition that there are only 1000, 2000, 3000 and 4000 labeled data in each class sample, and compares the classification results with CNN under the same conditions to prove the superiority of ByteSGAN in semi-supervised encrypted traffic classification.

\subsubsection{Performance result}\label{exp:config}
\begin{figure}[ht!]
	\centering\includegraphics[width=3.3 in]{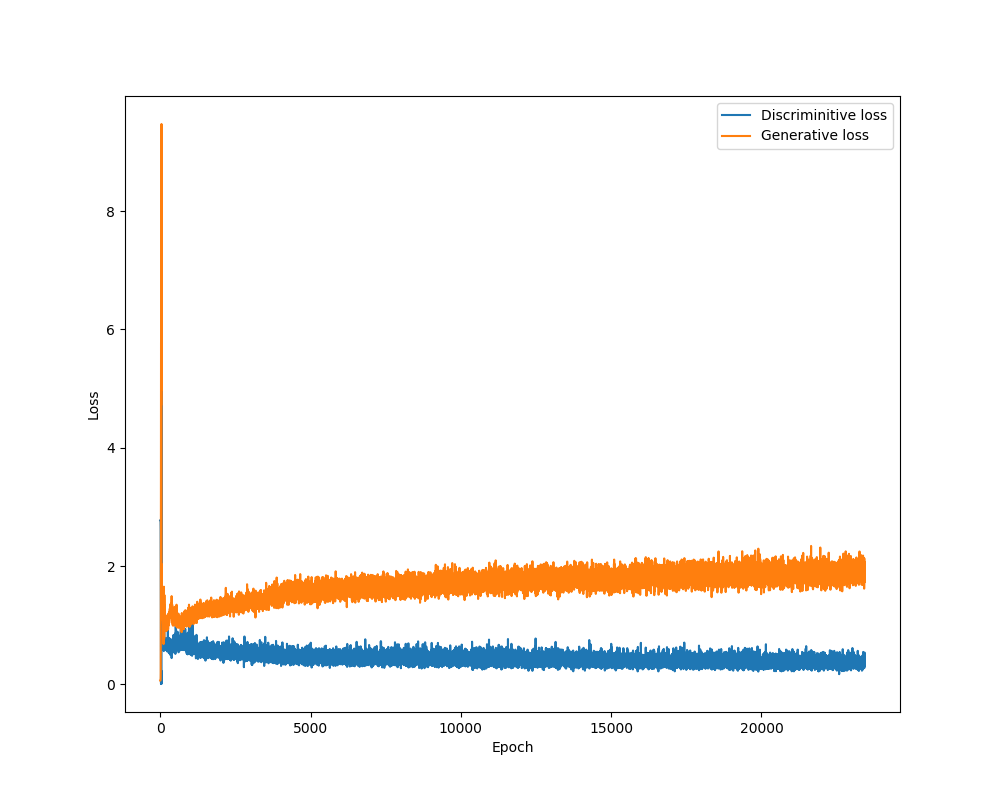} 
	\caption{ByteSGAN adversarial training loss}\label{fig:SGAN_loss} 
\end{figure}

Fig.~\ref{fig:SGAN_loss} shows the trend of the adversarial loss on the ByteSGAN discriminator and generator. The adversarial loss of the discriminator gradually decreases, and that of the generator gradually rises, and both of them keep stable around 2000 epoches.

\subsubsection{Classification result}\label{exp:classifier}
We employ ByteSGAN and CNN to perform classification experiments when the number of labeled samples is 1000, 2000, 3000, 4000, and the classification results are as follows:

\begin{table}[htbp]
	\caption{The second experimental results}  \label{tab:Experimental Two Results}
	\fontsize{6.5}{8}\selectfont  
	\begin{center}
		\begin{tabular}{c| c| c}
			\hline
			\textbf{Number of labeled samples (per application)}&{\textbf{ByteSGAN}} &{\textbf{CNN}} \\
			\hline
			1000		&92.15\%&88.25\%\cr  
			2000		&92.92\%&89.60\%\cr  
			3000		&93.10\%&92.40\%\cr  
			4000		&93.18\%&93.30\%\cr  
			\hline
		\end{tabular}
	\end{center}
\end{table}
Apparently, the experimental result shows that when the number of labeled samples is 1000, the classification accuracy of ByteSGAN is improved by about 5\% compared to CNN. When the number of labeled samples is 2000, the classification accuracy of ByteSGAN is improved by about 3\%. And when the number is 3000, ByteSGAN's classification accuracy is improved by less than 1\%. When the number of labeled samples is 4000, the classification accuracy is basically the same as that of CNN. It can be seen that when the number of labeled samples is insufficient, ByteSGAN can effectively improve the classification accuracy by using unlabeled samples and samples generated by ByteSGAN to perform semi-supervised learning.
The accuracy is still not a good reflection of the classification performance of ByteSGAN. Below we use the evaluation metrics to analyze the improvement of each application more comprehensively.
\begin{enumerate}
	\item Precision:
\begin{figure}[htbp]
	\centering
	\subfigure[1000 labeled samples]{
		\begin{minipage}[t]{0.5\linewidth}
			\centering
			\includegraphics[width=1.6in]{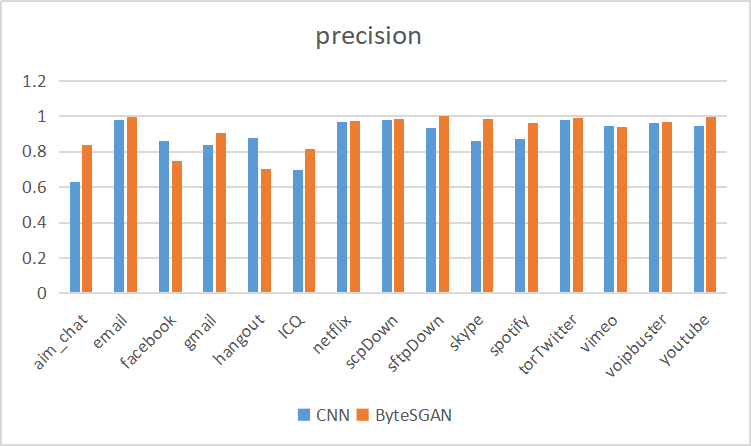}
		\end{minipage}%
	}%
	\subfigure[2000 labeled samples]{
		\begin{minipage}[t]{0.5\linewidth}
			\centering
			\includegraphics[width=1.7in]{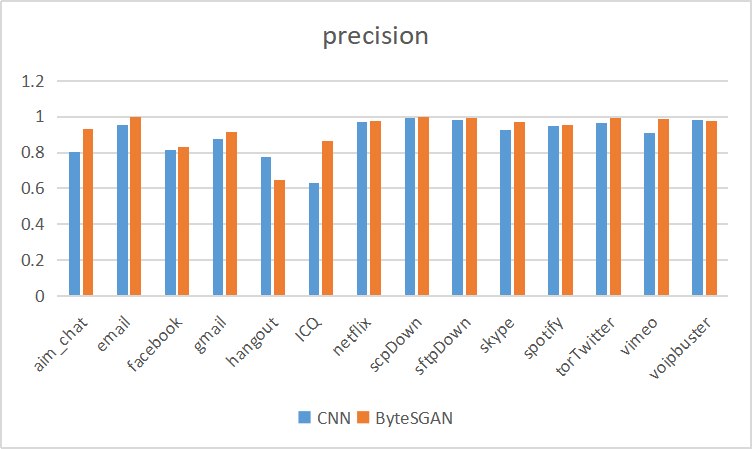}
		\end{minipage}%
	}%
	\\
	\centering
	\subfigure[3000 labeled samples]{
		\begin{minipage}[t]{0.5\linewidth}
			\centering
			\includegraphics[width=1.6in]{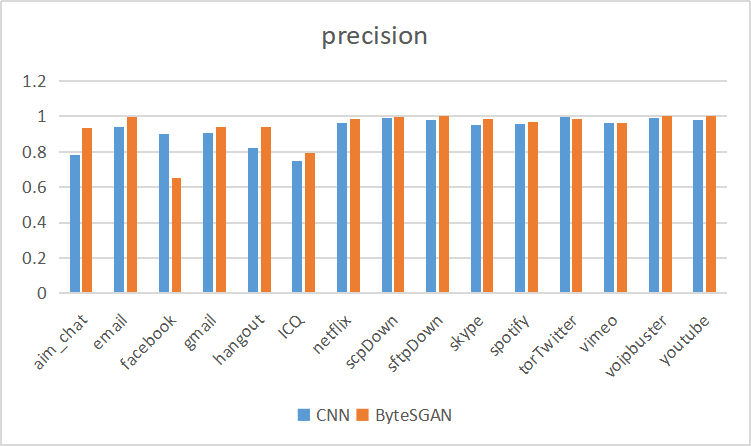}
		\end{minipage}%
	}%
	\subfigure[4000 labeled samples]{
		\begin{minipage}[t]{0.5\linewidth}
			\centering
			\includegraphics[width=1.7in]{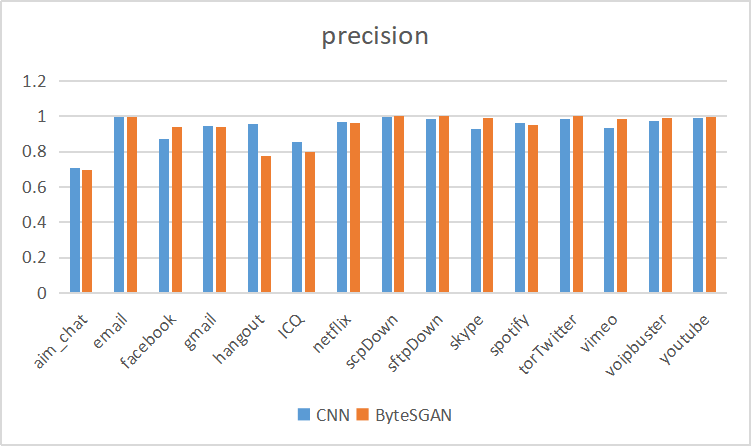}
		\end{minipage}%
	}%
	
	\caption{Comparison of precision}
	\label{fig:Comparison of precision indicators}
\end{figure}
As shown in Fig.~\ref{fig:Comparison of precision indicators}, except facebook and hangout, ByteSGAN's precision is basically better than that of CNN in 1000, 2000, and 3000 samples. Among them, the improvement of aimchat is higher and the improvement is close to 20\%. The increasing rate of email, gmail and ICQ is close to 5\% and other applications such as netflix, scpdown, skype, spotify, torTwitter, voipbuster, youtube also have a small increase.
	\item Recall:
\begin{figure}[htbp]
	\centering
	\subfigure[1000 labeled samples]{
		\begin{minipage}[t]{0.5\linewidth}
			\centering
			\includegraphics[width=1.6in]{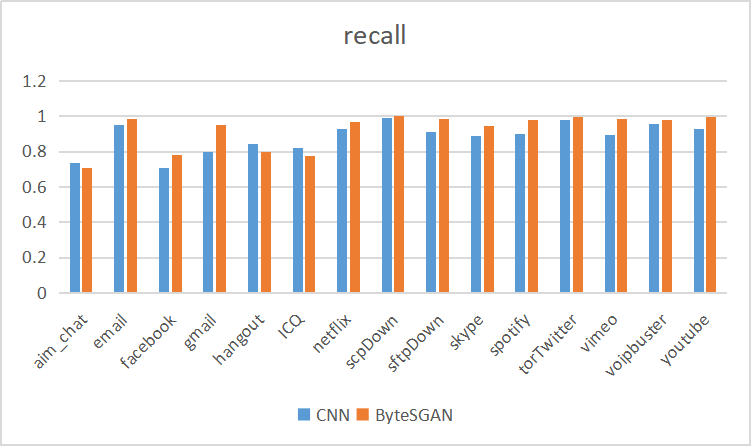}
		\end{minipage}%
	}%
	\subfigure[2000 labeled samples]{
		\begin{minipage}[t]{0.5\linewidth}
			\centering
			\includegraphics[width=1.7in]{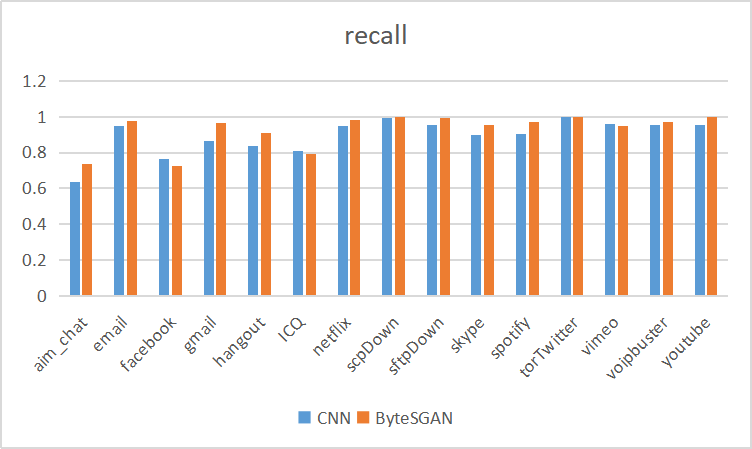}
		\end{minipage}%
	}%
	\\
	\centering
	\subfigure[3000 labeled samples]{
		\begin{minipage}[t]{0.5\linewidth}
			\centering
			\includegraphics[width=1.6in]{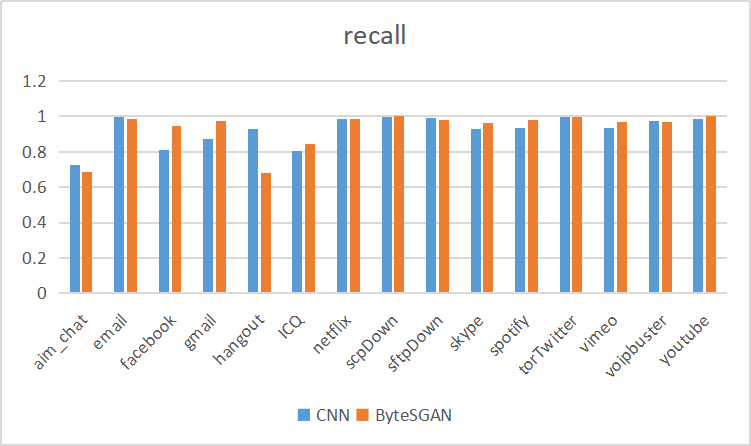}
		\end{minipage}%
	}%
	\subfigure[4000 labeled samples]{
		\begin{minipage}[t]{0.5\linewidth}
			\centering
			\includegraphics[width=1.7in]{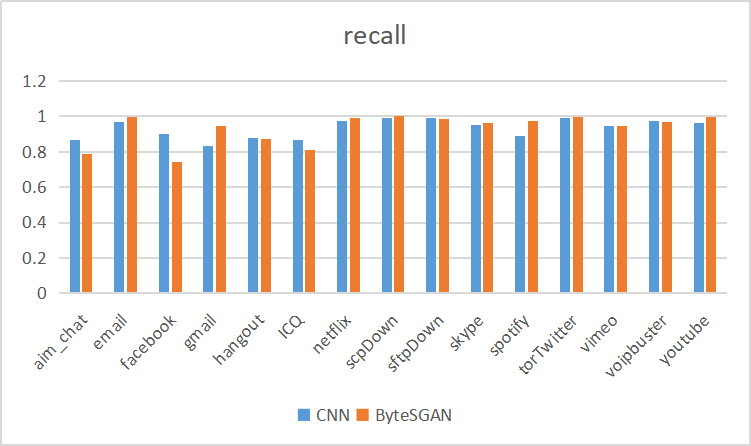}
		\end{minipage}%
	}%
	
	\caption{Comparison of recall}
	\label{fig:Comparison of recall indicators}
\end{figure}
As shown in Fig.~\ref{fig:Comparison of recall indicators}, in the experiment of 1000 samples, 2000 samples, and 3000 samples, SGAN has excellent recall for the 10 applications of gmail, netfliex, scpDown, sftpDown, skype, spotify, torTwitter, vimeo, voipbuster, and youtube. For CNN, the improvement of gmail is more significant, close to 20\%.

\item F1-score:
As shown in Fig.~\ref{fig:Comparison of recall indicators}, similar to the precision, except for the facebook and hangout applications, SGAN's f1-score index is basically better than CNN in 1000, 2000, and 3000 samples. Among them, aimchat has a higher improvement, which is an increase of about 5\% -10\%; email, gmail, ICQ have increased significantly. Other applications such as netflix, scpdown, skype, spotify, torTwitter, voipbuster, youtube also have a small improvement.
\begin{figure}[htbp]
	\centering
	\subfigure[1000 labeled samples]{
		\begin{minipage}[t]{0.5\linewidth}
			\centering
			\includegraphics[width=1.6in]{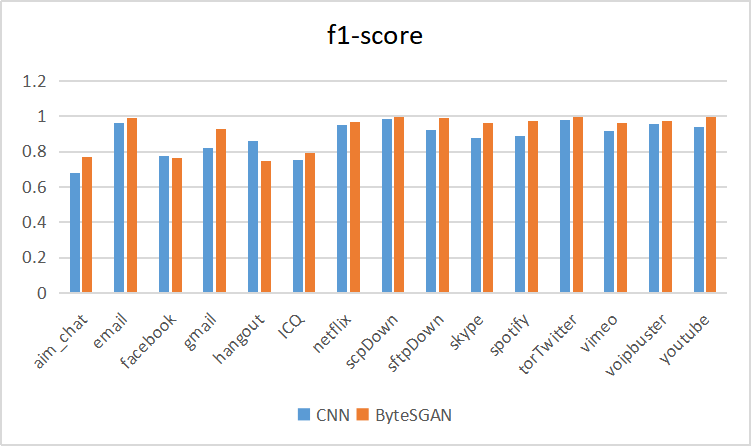}
		\end{minipage}%
	}%
	\subfigure[2000 labeled samples]{
		\begin{minipage}[t]{0.5\linewidth}
			\centering
			\includegraphics[width=1.7in]{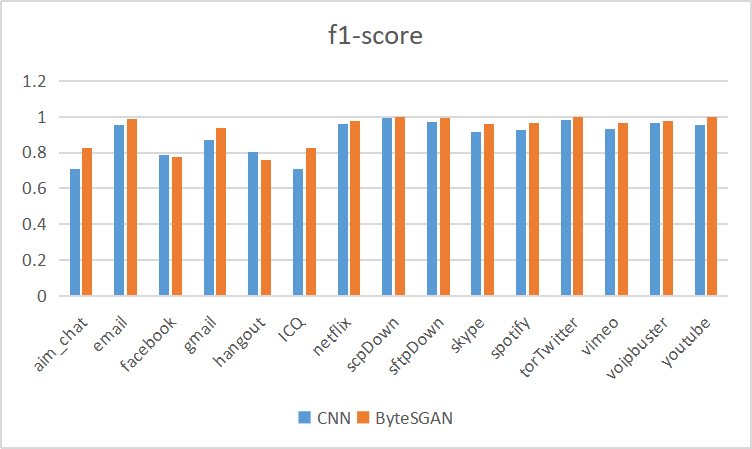}
		\end{minipage}%
	}%
	\\
	\centering
	\subfigure[3000 labeled samples]{
		\begin{minipage}[t]{0.5\linewidth}
			\centering
			\includegraphics[width=1.6in]{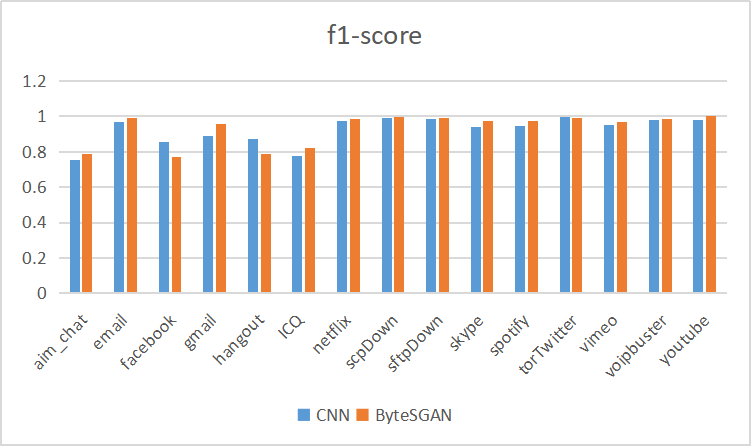}
		\end{minipage}%
	}%
	\subfigure[4000 labeled samples]{
		\begin{minipage}[t]{0.5\linewidth}
			\centering
			\includegraphics[width=1.7in]{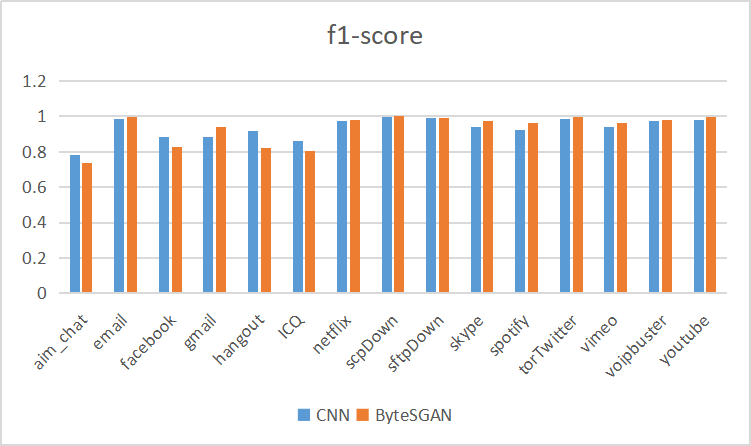}
		\end{minipage}%
	}%
	
	\caption{Comparison of f1-score}
	\label{fig:Comparison of f1-score indicators}
\end{figure}

\end{enumerate}

\section{Conclusion and Future work}\label{conclusion}
As we all know, capturing large labeled datasets is a cumbersome and time-consuming manual labor. Semi-Supervised learning is a desirable learning way to alleviate this problem. Motivated by this idea, we proposed a Generative Adversarial Network (GAN)-based Semi-Supervised Learning Encrypted Traffic Classification method called \emph{ByteSGAN} embedded in SDN Edge Gateway to achieve the goal of traffic classification in a fine-grained manner to further improve network resource utilization. ByteSGAN can only use a small number of labeled traffic samples and a large number of unlabeled samples to achieve a good performance of traffic classification by modifying the structure and loss function of the regular GAN discriminator network in a semi-supervised learning way. Based on public dataset 'ISCX2012 VPN-nonVPN', two experimental results show that the ByteSGAN can efficiently improve the performance of traffic classifier and outperform the other supervised learning method like CNN. In the future, we will further explore the potential of generative capability of GAN to improve the data imbalance problem for traffic classification.

\section*{Acknowledgment}
The paper is sponsored by National Natural Science Fundation (General Program) Grant 61972211, China, and National Key Research and Development Project Grant 2020YFB1804700, China.

\renewcommand\refname{Reference}
\bibliographystyle{IEEEtran}
\bibliography{IEEEfull,Reference}

\end{document}